# CLUSTER: A HIGH FREQUENCY H-MODE COUPLED CAVITY LINAC FOR LOW AND MEDIUM ENERGIES

Ugo Amaldi[a,b,*], Alessandro Citterio[a], Massimo Crescenti[a,†], Arianna Giuliacci[a], Cesare Tronci[a,§], Riccardo Zennaro[a]

[a] *TERA Foundation, Via Puccini 11, 28100 Novara, Italy*
[b] *University of Milano Bicocca, Milano, Italy*

**Abstract**

We propose an innovative linear accelerating structure, particularly suited for hadrontherapy applications. Its two main features are compactness and good power efficiency at low beam velocities: the first is achieved through a high working frequency and a consequent high accelerating gradient, the second is obtained by coupling several H-mode cavities together. The structure is called CLUSTER, which stands for "*Coupled-cavity Linac USing Transverse Electric Radial field*".
In order to compare the performance of this structure with other hadrontherapy linac designs involving high frequencies, a conceptual study has been performed for an operating frequency of 3 GHz. Moreover a proof of principle has been obtained through RF measurements on a prototype operating at 1 GHz.
An accelerator complex using a CLUSTER linac is also considered for protontherapy purposes. The whole complex is called cyclinac and is composed of a commercial cyclotron injecting the beam in a high-frequency linac.

*Keywords*: cyclinac, linac, hadrontherapy, H-cavity, low-$\beta$ structure, coupled structure.

---

[*] Corresponding author: *CERN, CH-1211 Geneva 23, Switzerland, Tel +41 22 76 73876, Fax +41 22 76 79740*, e-mail *Ugo.Amaldi@cern.ch*.
[†] Now at European *Patent Office, Patentlaan 2, 2288 EE Rijswijk, The Netherlands*.
[§] Currently also at *Department of Mathematics, Imperial College London, London SW7 2AZ, UK*.

# 1. THE CYCLINAC APPROACH TO HADRONTHERAPY

This paper describes a novel type of drift tube linac which is optimally suited to accelerate slow nuclear particles ($\beta \leq 0.2$) with high field gradients. Since the structure (named *CLUSTER*), has several potential applications, in the rest of the paper no particular use is described. However this first section is devoted to explain the framework in which the idea was born.

In modern hadrontherapy, solid tumours are controlled by irradiating them with beams of charged hadrons, in particular protons and carbon ions [1]. The rationale of protontherapy is the dose depth distribution characterized by the Bragg peak allowing a dose delivery which spares healthy tissues much better than 10 MeV photons (X rays) [2]. Clinically protons have the same effect as X-rays, so that the knowledge accumulated with conventional therapy can be immediately transferred to this technique. A carbon ion leaves about twenty-four times the energy in each cell compared to a proton of the same range and produces qualitatively different effects. Indeed, carbon ion therapy targets successfully radio resistant tumours, i.e. the slowly growing tumours that are insensitive to both X-rays and protons [3, 4, 5].

In 2006, hospital-based centres for deep protontherapy are treating patients in the United States (3), in Japan (4) and in China (1). Other hospital-based centres for protons are under construction in China, Italy, Germany, Switzerland, South Korea and United States. In 2007, two other European centres will be completed. The first, called HIT, was designed by GSI and is located in Heidelberg, Germany, the second one, CNAO, has been designed by the TERA Foundation and is currently being built in Pavia, Italy. Since the beginning of the century hadrontherapy is developing very rapidly, so that six companies offer turn-key facilities, which are based on either synchrotrons or cyclotrons. Only in the last few years superconducting cyclotrons for carbon ion therapy have been designed [6].

As an alternative to cyclotrons and synchrotrons, the TERA group has introduced a novel accelerator system, called "cyclinac", which is composed of a low to medium energy cyclotron and a high-frequency, high-gradient linac. The usual beam lines and gantries are located at the end of the linac. It can been shown that the time and intensity structure of a cyclinac is better suited to the active dose distribution approach, called "spot scanning" and developed at PSI [7] than those produced by cyclotrons and synchrotrons. A cyclinac is intended to be the heart of IDRA, the "Institute for Diagnostic and RAdiotherapy", which is a multipurpose facility for the production of both radiopharmaceuticals for diagnostic and therapy, and high-energy protons for the radiotherapy of superficial and deep-seated tumours (Fig. 1).

In IDRA one of the 30 MeV cyclotron beam lines injects protons in a linac booster (LIBO), which is a Side Coupled Linac (SCL) running at 3 GHz, the same frequency adopted by more than 8000 electron linacs used in the world for conventional radiotherapy. The idea of the cyclinac goes back to 1993 [8]. The $f = 3$ GHz choice leads to a relatively short linac, since the achievable gradient is roughly proportional to $f^{1/2}$, and gives access to a wide market of the necessary pulsed klystrons and accessories. In 1998 a prototype of LIBO was built and power tested, with an accelerating gradient of 27.5 MV/m [9, 10]. The acceleration tests performed at the superconducting cyclotron of the INFN Laboratorio Nazionale del Sud (LNS) were also very satisfactory [11,12].

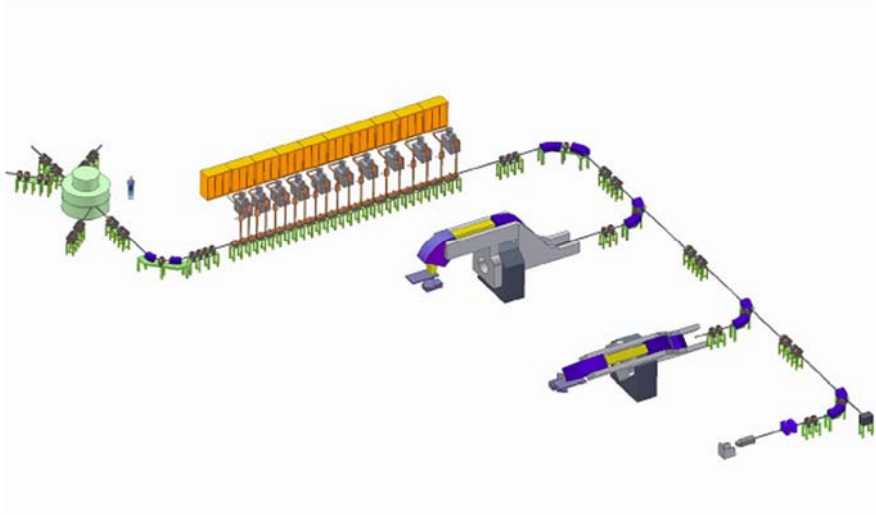

*Figure 1. Artistic view of the IDRA complex made of a high-current 30 MeV commercial cyclotron and a linear accelerator of the LIBO type.*

In the version of LIBO designed for IDRA, the protons enter with an energy of 30 MeV/u, i.e. a velocity $\beta = 0.25$. Lower energies cannot be accepted because the shunt impedance of a SCL structure drastically decreases for $\beta \leq 0.2$. This is the first reason for the study of new accelerating structures that are more efficient at low energies. The second reason concerns the design of cyclinacs for carbon ion therapy. For these applications the existing low cost commercial cyclotrons have maximum $C^{+6}$ energies of the order of 10 MeV/u, that is too low for direct injection into an SCL structure.

These motivations lead to the CLUSTER concept described in the rest of the paper, which is however very interesting for many other applications and thus stands by itself. In Section 2 we describe the structure and explain its physical principles. In the Section 3 the scheme of both the coupling cavities and the end cells is presented. Section 4 contains a detailed presentation of how each single component has been studied, adopting a reference frequency of 3 GHz. A simple circuit scheme is also described. The experimental results of a 1 GHz prototype are presented in Section 5, while Section 6 discusses a design of a cyclinac employing a 1.5 GHz CLUSTER structure.

## 2. CLUSTER, A NOVEL LINAC STRUCTURE

The need for a high shunt impedance in the low-medium $\beta$ range (0.05-0.4) leads to the choice of H-mode accelerating cavities, also called TE (Transverse Electric) cavities since the electric field is naturally transverse to the axis. These structures have been studied since the 50's [13, 14] and are nowadays successfully used, for example in Unilinac at GSI [15] and in Linac3 at CERN [16], both working at low frequencies (between 100 and 200 MHz).

H-mode cavities are drift tube cavities operating in the $H_{n1(0)}$ mode, where the index $n$ is usually 1 (IH cavities; already existing) or 2 (CH cavities, under development). These cavities are very attractive because of the high shunt impedance for low $\beta$'s. This feature is due to the fact that the electric field is mostly concentrated close to the axis in the accelerating gaps, where it is made parallel to the axis by the metallic drift tubes. Moreover, they are π-mode structures, i.e. the RF accelerating field is phase shifted by 180° between successive gaps. Such structures allow higher average gradients, which are further increased in the present case because of the high frequency (above 1 GHz).

Large values of the operating frequency imply two main problems: the location of the magnetic quadrupoles for beam focusing and the field sensitivity to mechanical imperfections (i.e. frequency perturbations), commonly referred to as "field instability". These problems are best discussed with reference to the CLUSTER scheme of Fig. 2, which is composed of a sequence of four accelerating sections (tanks) alternating with three coupling cells.

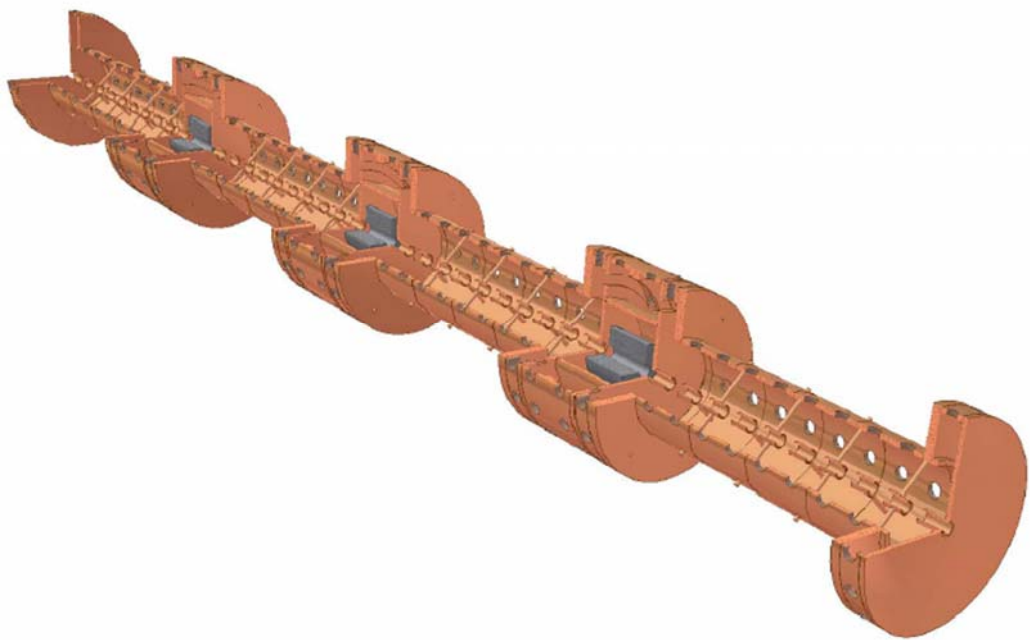

***Figure 2.*** *A module of CLUSTER, the "Coupled-cavity Linac USing Transverse Electric Radial" field. The accelerating tank consists of a sequence of gaps and drifts whose length is constant in a tank (constant β).*

The accelerating sections operate in the π/2 beam mode and contain *m* gaps. They are connected by coupling sections consisting of coaxial cylindrical cavities, which allow the location of Permanent Magnet Quadrupoles (PMQ's), thus solving the first problem.

The second problem (field instability) is related to the frequencies of the perturbing resonances. Some quantitative considerations may be made by the analogy between H-mode structures and the four-vane RFQ. Indeed, in the latter case, the frequency $\omega_n$ of the *n*th mode satisfies the following relationship [17]:

$$\left(\frac{\omega_n}{\omega_0}\right)^2 = \left(\frac{n\cdot\lambda}{L}\right)^2 + 1 \tag{1}$$

where $\omega_0$ is the fundamental frequency corresponding to the working mode, $\lambda$ is the RF wavelength and $L$ is the length of the RFQ. Eq. (1) can be obtained from an equivalent circuit approach where the capacitance is uniformly distributed along the cavity, and it is a very good approximation of the dispersion curve especially for small values of $n$. In the case of IH or CH cavities, where the capacitance is concentrated in a discrete number of gaps ($m$), Eq. (1) needs to be modified accordingly. Indeed, since in an H-structure the length $L$ is proportional to the number $m$ of accelerating gaps, one may define a positive quantity $\theta$ such that

$$\left(\frac{\omega_n}{\omega_0}\right)^2 = \theta \cdot \left(\frac{n}{m}\right)^2 + 1 \qquad (2)$$

where the geometric factor $\theta$ has to be evaluated on a case by case basis. This formula is valid for both IH and CH structures and has been verified in the case of CLUSTER.

It has to be noticed that $\theta \neq (2/\beta)^2$ in contrast with what would be obtained from Eq. (1) through the substitution $L=m\beta\lambda/2$. Rather, $\theta$ depends on the geometry of the cavity. For a fixed $\beta$, the number of gaps for unit length increases with the frequency and so does the field instability because, according to Eq. (2), the frequencies $\omega_n$ of the higher order modes get closer to $\omega_0$. This puts severe limitations on the length of such high-frequency accelerating cavities since, in most practical cases, the number of cavities would become too large for a given energy step.

Solutions to this problem have been proposed in the past for TM cavities of the type CCDTL [18] or SCDTL [19]. In particular, the second design refers to a 3 GHz structure intended for protontherapy. Both these solutions foresee a coupled-cavity structure operating in the π/2 mode: accelerating tanks are alternated to coupling cavities so that the field phase shift between them is π/2 and the electric field is zero in the coupling cavities. This configuration allows the field to keep the right configuration, because it is less sensitive to the unavoidable frequency perturbations [20]. Moreover PMQ's could be placed outside the accelerating tanks, instead of inside the drift tubes as is usually done in Alvarez linacs. The shunt impedance is then higher than in conventional Alvarez structures, but still smaller, at low velocities, than that achievable with the CLUSTER TE structure design.

## 3. COUPLING CELLS AND END CELLS

The coaxial shape of the coupling cell of Fig. 3 has been chosen, having in mind the simplicity of machining and easiness of PMQ insertion. The TEM mode has a simple field pattern in coaxial cavities.

For the coupling cell, the RF wave-particle synchronism requires a phase advance of 2π between the centers of the two accelerating gaps beside the coupling cell, so that the following equation has to be satisfied:

$$L = n\beta\lambda_0$$

where $L$ is the distance between the centres of the consecutive gaps and $n$ is positive integer. On the other hand, a simple coaxial cavity of length $l$ and resonating in a TEM mode satisfies the dispersion equation:

$$2\nu_p l = pc$$

where $\nu_p$ is the $p^{th}$ TEM frequency and $c$ is the speed of light. These two constraints are both satisfied only in particular cases and a simple tuning solution had to be found in order to allow the design of compact coupling cells for the whole $\beta$ range.

The insertion of an annular capacitive tuner in the cavity centre is a satisfactory solution to the problem. The geometric dimensions have been chosen in order to both allow PMQ positioning and also have a frequency close to that of the accelerating tanks, according to the theory of resonant coupling [20].

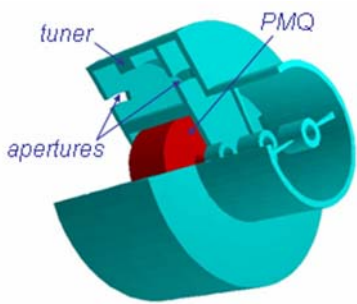

*Figure 3.*   *A PMQ inserted in a coupling cell.*

An important advantage of this configuration is the possibility of producing the coupling cell transversally split in two identical parts, which are successively flanged together. The PMQ can be easily inserted inside the cell before joining the two parts. In case of failure or bad PMQ positioning, the cell can be re-opened and the PMQ easily replaced or repositioned. In fact, no tight surface contact (i.e. brazing) is required due to the field pattern of the $TEM_{011}$ mode: at the centre of the coaxial cavity the RF currents are zero on both the inner and outer surfaces. Moreover, this solution greatly simplifies the brazing procedure of the entire structure: the brazing would be limited to several short mono-blocs, making possible the use of short vacuum furnaces.

As indicated in Fig. 3, the coupling is obtained through four apertures (slots) on each side of the coupling cell. The coupling is magnetic (i.e. with zero electric field in that region) and the slots are shaped to guarantee the correct coupling for the required field stability. The coupling factor is determined by the shape and particularly by the length of the slots; it is roughly proportional to $L^3$, where L is the slot length [21]. A convenient value for the coupling is a compromise between the power efficiency, which decreases with the coupling, and the field stability which requires a large pass-band for the dispersion curve, i.e. a large coupling.

The configuration of the end-cells in each tank has been studied to allow suitable return paths for the magnetic field lines, in order to keep the accelerating field as uniform as possible in each gap. The appropriate field pattern is obtained by increasing the diameter of the end-cell, a simple solution for machining. A view of the end of a tank with an end-cell is shown in Fig. 4.

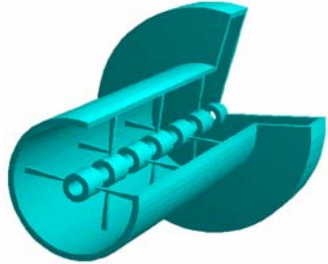

*Figure 4.*     *Accelerating gaps and end-cell.*

## 4. RF COMPUTATIONS AND THE DESIGN OF CLUSTER

The RF design has been performed using the MAFIA code [22] and CST Microwave Studio (MWS) [23].

### 4.1 Choice of the accelerating mode

The first step in the design is the choice of the working mode. To this purpose simple shapes have been considered (Fig. 5).

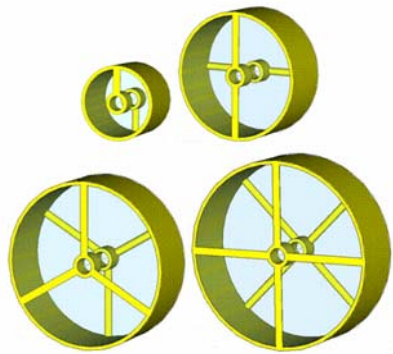

*Figure 5.*     *The four structures used in the comparison. The number n of stems per drift cell (n = 1,2,3,4) is the index of the oscillating mode $H_{n1(0)}$.*

The candidates are the $H_{n1(0)}$ modes ($n$ = 1,2,3,4) obtained with the geometries of Fig. 5, for which the basic RF parameters, namely the quality factor Q and the effective shunt impedance $ZT^2$, have been compared at the nominal frequency $f$ = 3 GHz.

The use of frequencies which are an order of magnitude higher than those normally employed for this kind of cavities makes it impossible to simply scale down the overall dimensions. At low frequencies, the ratio between the cavity diameter ($D_c$) and the drift tube diameter ($D_d$) is very large: for the GSI-HLI IH linac $D_c/D_d \approx 22$ [24]. At high frequencies, while $D_c$ is scaled down roughly as $1/f$, this ratio must be drastically reduced to provide a reasonable beam acceptance of the linac. The comparative study of four cavities is summarized in Fig. 6, which considers $H_{n1(0)}$ modes with $n = 1,2,3,4$, and points to small value of $D_c/D_d$ especially for $n = 1$ (IH cavity).

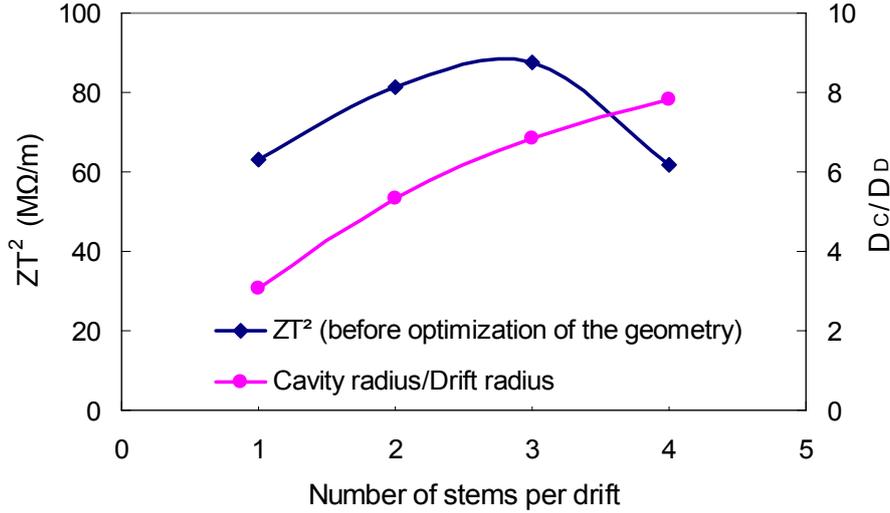

**Figure 6.** *Comparison of four different 3 GHz structures for $\beta = 0.28$ and $D_d = 6$ mm.*

For a TE mode, a small value of $D_c/D_d$ implies a strong component of the radial electric field between the drift tubes and the outer tank and thus a low shunt impedance, especially for the $n = 1$ case as shown on the left scale of Fig. 6. The figure shows that $ZT^2$ grows with $D_c/D_d$, with the exception of the fourth mode, for which the large losses on the stems/drifts reduce the value of $ZT^2$. The best results in terms of effective shunt impedance are obtained with the $n = 3$ mode, hereafter called 3H cavity, but the difference with respect to the more conventional CH mode ($n = 2$) is small. This latter solution has been adopted as reference structure in the CLUSTER design. The 3H structure might be an interesting solution for future work. The comparison of Fig. 6 is made at a particle velocity $\beta = 0.28$ (E = 40 MeV/u).

In Fig. 7 the shunt impedances of the IH and CH cavities are compared for different beam energies. The curves of Fig. 7 have been obtained by keeping constant both the working frequency (f = 3 GHz) and the drift over gap length ratio as well as by adjusting the external diameter ($D_c$). In both cases the shunt impedance decreases with the beam energy, as expected for all H-mode cavities. In the IH case, the reduction is more evident and makes the IH cavity inadequate for medium energy linacs. In the basic equivalent circuit model a single cell can be represented by a single "LCR" loop. The value of the frequency, Q and Z obtained by MWS computations allow to find the values of L (inductance), C (capacitance) and R (resistance). The capacitance C can be represented as the sum of two parallel components: the capacitance $C_g$ in the gap and the capacitance $C_r$ between the drifts and the outer tank. In both cases, the reduction of the shunt impedance for large $\beta$ is related to two factors. The first is the increased losses, which are due to the longer cavities. The second is the increased contribution of the capacitance $C_r$ to C: indeed, $C_r$ depends on the radial electric field between the outer tank and the drifts, whose length increases with $β$. As shown in Fig. 8, in the case of IH cavities (small radius) $C_r$ grows rapidly with $β$. This is not the case for CH cavities, which have a larger radius: this explains the better performance of the CH structure at medium-high values of $β$.

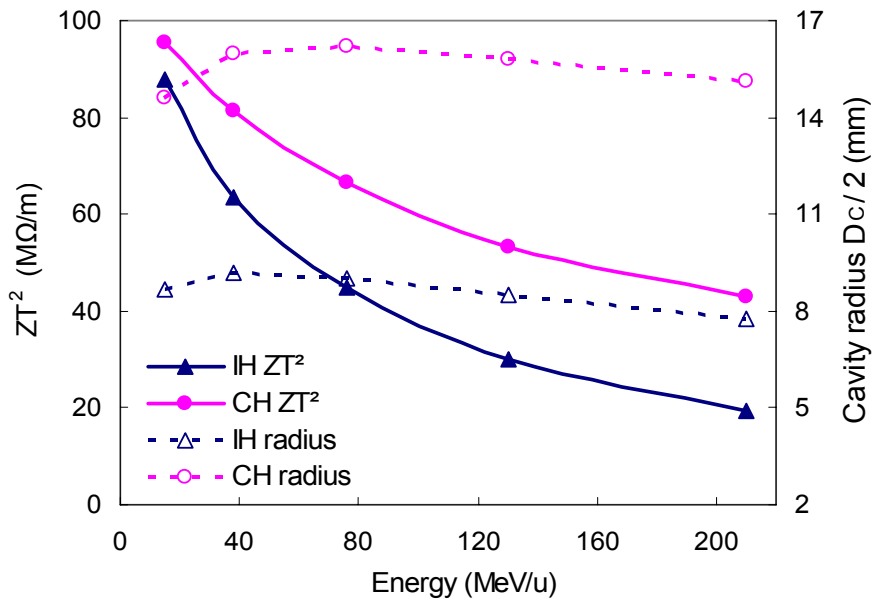

***Figure 7.***   *IH-CH comparison of $ZT^2$ and cavity radius for different cavity radii ($D_d$ = 6 mm).*

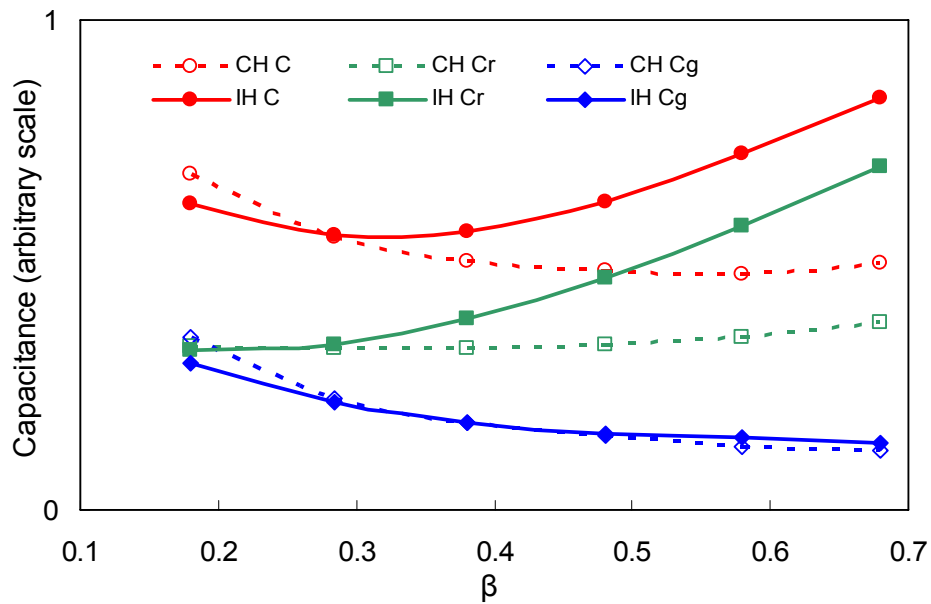

***Figure 8.***   *The capacitances $C_g$, $C_r$, $C = C_g+C_r$, as a function of the particle velocity $β$.*

The study in Fig. 8 was performed with cylindrical stems (Fig. 5). Different geometries have been evaluated, also with thermo-mechanical studies to be reported in a future work. The preferred geometry has conic stems as shown in Fig. 9.

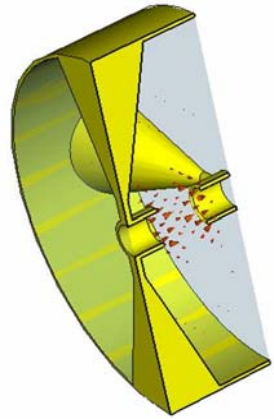

*Figure 9.*     *Optimized shape of the stems.*

The modification of the shape of the stems, from cylindrical to conical, increases the value of the shunt impedance. The use of conic stems, which have a larger volume, implies a larger cavity radius to keep the same resonant frequency. The result is a lower radial capacitance (Cr) and consequently larger shunt impedance. The energy dependence of the shunt impedance is shown in Fig. 10.

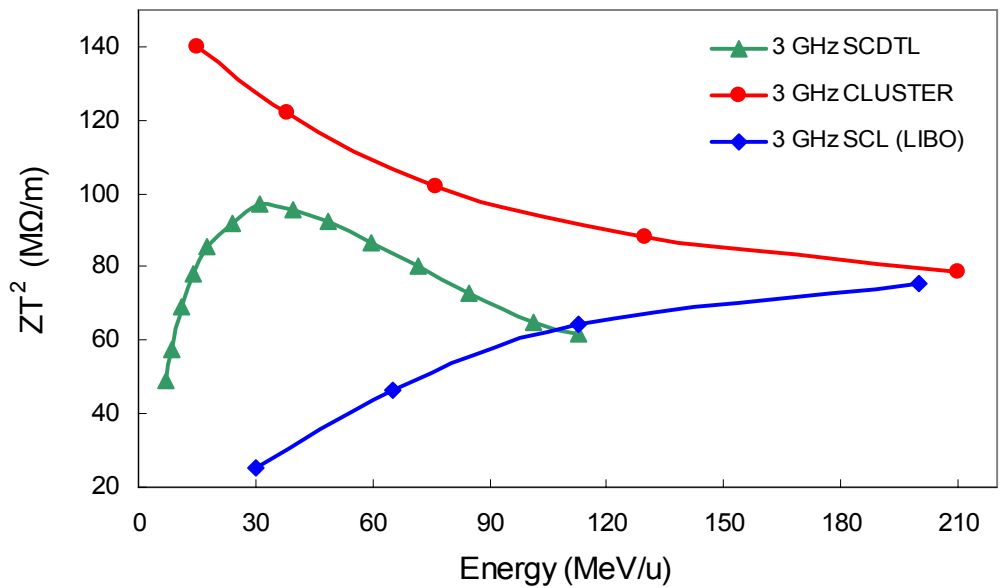

*Figure 10.*     *Effective shunt impedance of three 3 GHz linacs conceived for hadrontherapy.*

In a very large $\beta$ range (0.1-0.6) it turns out that CLUSTER has a higher effective shunt impedance $ZT^2$ than those of the other two structures designed for protontherapy: LIBO [25] and the SCDTL, developed by ENEA for the TOP project [26].

It should be noticed that, in Fig. 10, the beam hole radii are 2.0 mm, 2.5 mm and 4.0 mm respectively for CLUSTER, SCDTL and LIBO. The reduction of the LIBO beam hole to 2.0 mm would improve its shunt impedance by nearly 35% for low $\beta$; still CLUSTER is more efficient up to 140 MeV/u. A comparison of the three structures for a beam hole radius of 2.5 mm at the energy of 60 MeV/u is reported in Table 1.

*Table 1* *Comparison of the three structures.*

| ACCELERATING STRUCTURES | $ZT^2$ (M$\Omega$/m) |
|---|---|
| 3 GHz SCL (LIBO) | 55.4 |
| 3 GHz CLUSTER | 92.4 |
| 3 GHz SCL SCDTL | 86.7 |

CLUSTER is the most efficient structure also for the same beam hole radius as shown in Table 1.

### 4.2 The end cells

A noticeable result concerns the field behaviour in the end-cells, since the field pattern has to change radically from the $TE_{21(0)}$ mode into a $TM_{01(0)}$ mode. This pattern is completely different from the one characterizing the termination of a four-vane RFQ, where the angular component of the magnetic field has four zeros, corresponding to the four vanes. In a crossbar structure, drift tubes and stems compose discrete units. In the end cell, the last drift tube and its corresponding couple of stems break the quadrupole symmetry. This implies a $TM_{01(0)}$ mode configuration with a rotational magnetic field, whose angular component has no zeros although it is slightly modulated in amplitude.

Fig. 11 shows the field pattern in a transversal section of an end-cell: this is similar to a $TM_{01(0)}$ mode (see the arrow plot at the top) with an angular modulation of the amplitude (see magnetic energy density contour plot at the bottom). Notice that the largest magnetic field lies in correspondence with the vertical stems on the penultimate drift tube.

This pattern allows a simple magnetic coupling between the tanks and the coupling cell, since the situation is identical to the case of the coupling between two TM cavities having the same axis.

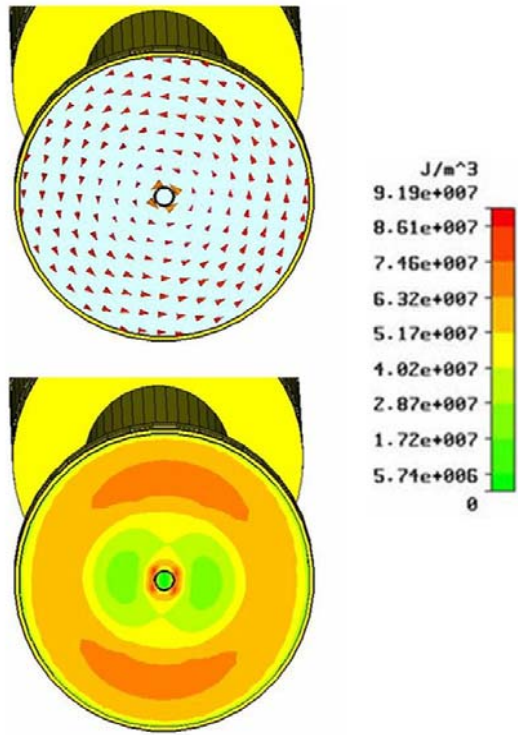

*Figure 11.* *Magnetic configuration in the end cell. The top figure is the arrow plot of the magnetic field. At the bottom is presented the contour-plot of the magnetic energy density in the middle plane section of the coupling cell.*

**4.3 Dispersion curve**

By using MWS, the dispersion curve for a full module of 36 accelerating gaps was computed. Three different configurations have been compared: a module composed of 3 accelerating sections with 12 gaps each, a module composed of 4 accelerating sections with 9 gaps each, and a module of 6 accelerating sections with 6 gaps each.

Each mode is characterized by its own phase advance between an accelerating section and a coupling cell. The amount of resonant modes, considered as composing the dispersion curve, is equivalent to the total number of oscillators in the structure. For the first case it corresponds to 5 (3 accelerating section and two coupling cells), in the second case to 7 and in the third case to 11.

As shown in Fig. 12, the frequency bandwidth grows with the number of sections; a large number of sections implies shorter accelerating tanks (fewer number of gaps) and consequently less stored energy per tank, which originates from a larger coupling.

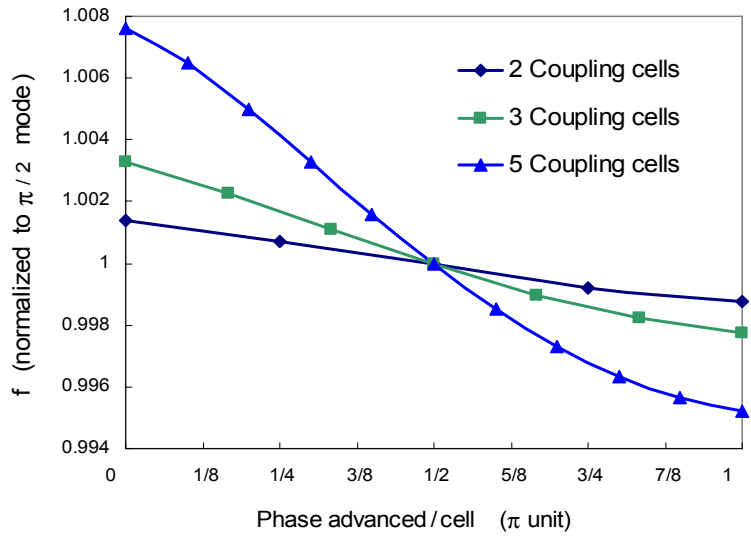

*Figure 12.    Dispersion curves for three different configurations of CLUSTER, computed with MWS.*

### 4.4 Equivalent circuit

The discrete model of coupled oscillators, based on a pure bi-periodic chain of oscillators, is not accurate enough to represent multi-gaps sections and can be applied to CLUSTER only in the case of a large number of short coupled sections. Many attempts with conventional tools like DISPER [27] have been performed in order to obtain a suitable fit to the dispersion curve calculated using the theory of coupled resonators. Nevertheless unsatisfactory results have been obtained, even for the structure with five coupling cells.

A more accurate study of the different resonances and of the dispersion curve can be performed using a different circuit approach. The analogy between RF cavities and coupled oscillators is a well known tool and, in the case of particle accelerators, this analogy leads to useful quantitative estimates.

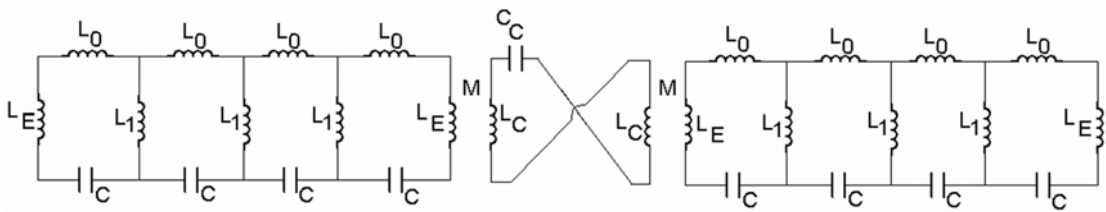

*Figure 13.    Example of a circuit with a single coupling cell*

A circuit model has been developed in order to reproduce the RF field behaviour in CLUSTER, starting from the definition of the coupling elements among the neighbouring gaps, each one conventionally modelled by LC loops (Fig. 13) [28].

Differently from the case of drift tube Alvarez cavities, where the coupling element is simulated by a capacitor, here, following Lee [29], the coupling is better simulated by the insertion of an inductor $L_1$ between the two neighbouring LC loops representing the oscillators. Moreover the total inductance in the terminal loops (end cells) has been represented by the series of two different inductors ($L_0+L_E$). The numerical value of $L_E$ is chosen in order to reproduce the right field modulation, with the same amplitude in each gap. In agreement with [30], one obtains $L_E = 2L_1$. On the other hand, a suitable scheme for the coupling cell is a mutual inductance. The current inversion at the middle of the cell is reproduced by bending the LC loop, as shown in Fig. 13.

A simple MATLAB program has been used to compare the behaviors of the RF structure and the chain of oscillators. The program is based on the matrix which characterizes the eigenvalue problem obtained by applying Kirchoff's laws to the circuit. The input data are the frequencies of the coupling cell and of the accelerating tank, the coupling value, the numbers of accelerating tanks and of gaps per tank, and also a non-dimensional parameter related to the stored energy in the coupling cell.

The results, presented in Fig. 14 and Fig. 15, are in perfect agreement with MWS simulations both for the frequencies (eigenvalues) and the field modulation (eigenvectors).

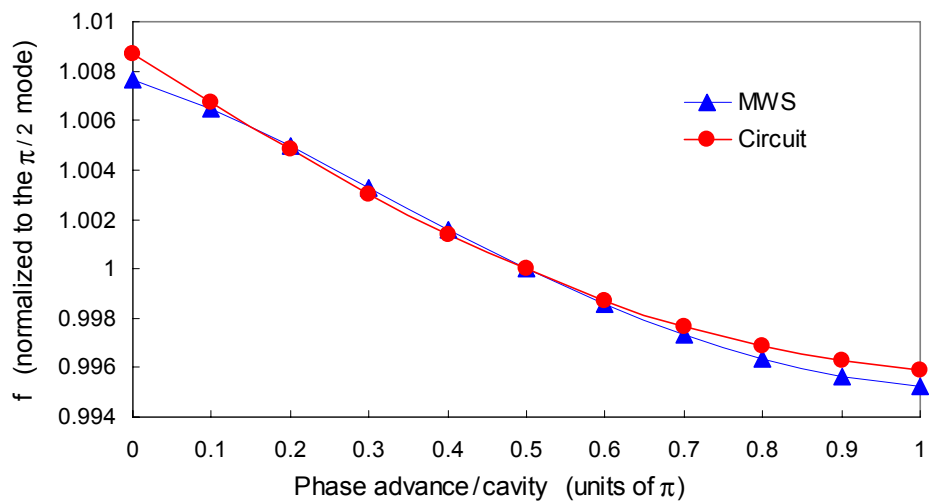

*Figure 14. Dispersion curves for a system of 6 accelerating tanks (6 gaps/tank) and 5 coupling cells. The dispersion curve obtained with the circuital approach fits very well the one obtained with MWS, especially for the $\pi/2$ working mode and the neighboring modes.*

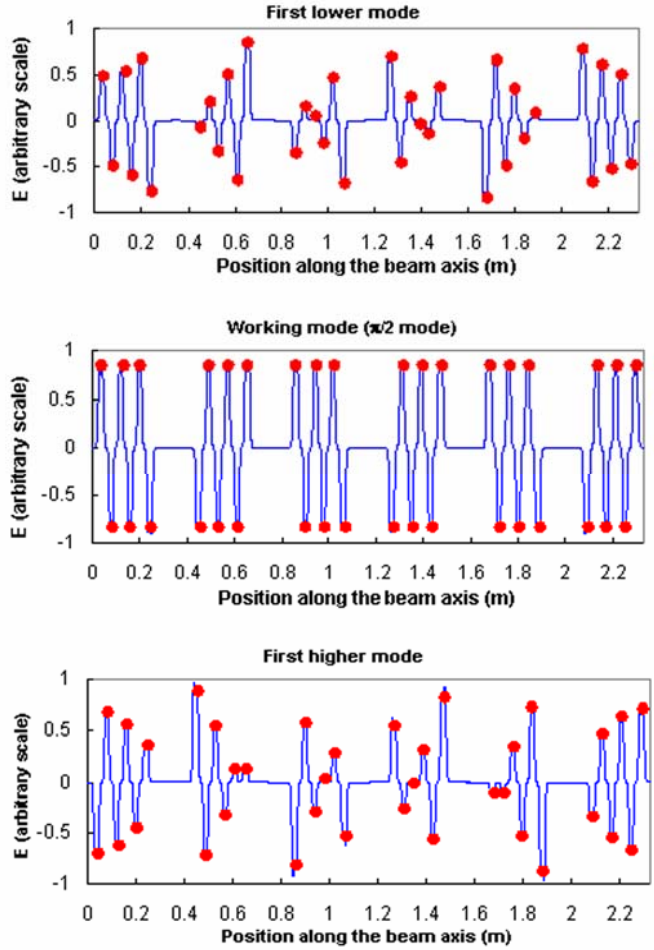

*Figure 15.* *Examples of the axial electric field pattern (eingenvectors) for the working mode and the two neighbour modes obtained with MWS (continuous line) and the MATLAB description of the equivalent circuit (points).*

## 5. CONSTRUCTION AND RF TESTS OF THE CLUSTER PROTOTYPE

A CLUSTER model has been built to test the RF design of the structure in terms of field behaviour and mode analysis. A frequency $f = 1$ GHz was chosen to have both a longitudinally compact structure and a certain simplicity in the design and the measurements. The results can be translated to future 1.5 or 3 GHz designs by using well known scaling laws [31, 32]. In the design, the $\beta$ value was kept constant since particle acceleration is out of the present scope. The value $\beta = 0.25$ was chosen, which corresponds to about 30 MeV protons.

The 1 GHz model has a modular geometry and is composed, in its full configuration, of two accelerating sections and a central coupling cell, as shown in Fig. 16. The beam hole diameter has been fixed at 12 mm, the drift tube thickness has been kept to 1.5 mm because this is a suitable value for the mechanical construction. The tank radius is 50 mm and the gap length (18 mm) is a consequence of the frequency choice. The total length of the structure in the full configuration is 1.1 m.

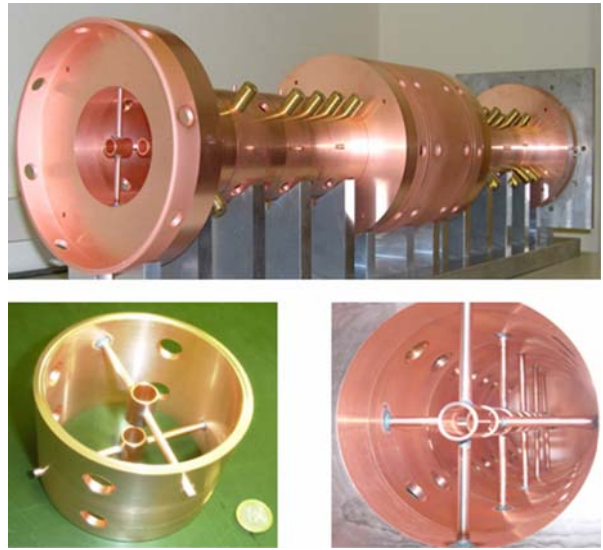

*Figure 16.    The prototype for low power RF measurements.*

The central coupling cell can be removed to have a single accelerating section, while the number of gaps can be varied at will. Four tuners per each gap (one per quadrant) are used to tune the structure and to study the accelerating field pattern in presence of perturbations. Screw tuners are also placed in the end-cells and coupling cells. The CLUSTER structure is designed so that the necessary focusing quadrupoles can be easily allocated.

The central ring in the coupling cell can be machined to tune the frequency of the system. The main mechanical tolerances (±0.2 mm) are rather low for a relatively high frequency structure.

Several measurements have been performed both in the frequency domain (dispersion curve) and in the time domain (bead pulling). As shown in Fig. 17, the validity of Eq. (1) is confirmed by the measurements of the frequencies for the different modes ($n$), with respect to the number of gaps ($m$). These measurements were made on the basic configuration, composed of a single accelerating section.

The fit presented in the Fig. 17, for the $\theta$ parameter in Eq. (2),. gives $\theta \cong 1.18$. The value of $\theta$ depends on the geometry of the structure and particularly on the β of the design. The values of $\theta$ for structures at different β but with the same frequency (3 GHz) and with identical geometry of the conic stems have been computed; see Fig. 18. The value of $\theta$ is proportional to 1/β which means that the problem of the RF stability is more severe for larger β structures.

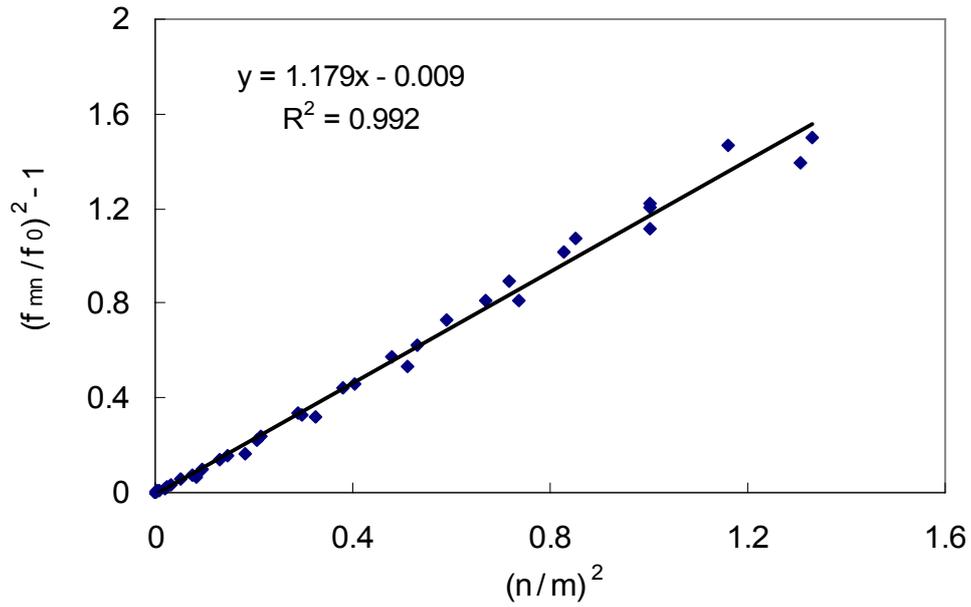

***Figure 17***.   *Linear fitting of Eq. (1) to $(n/m)^2$.*

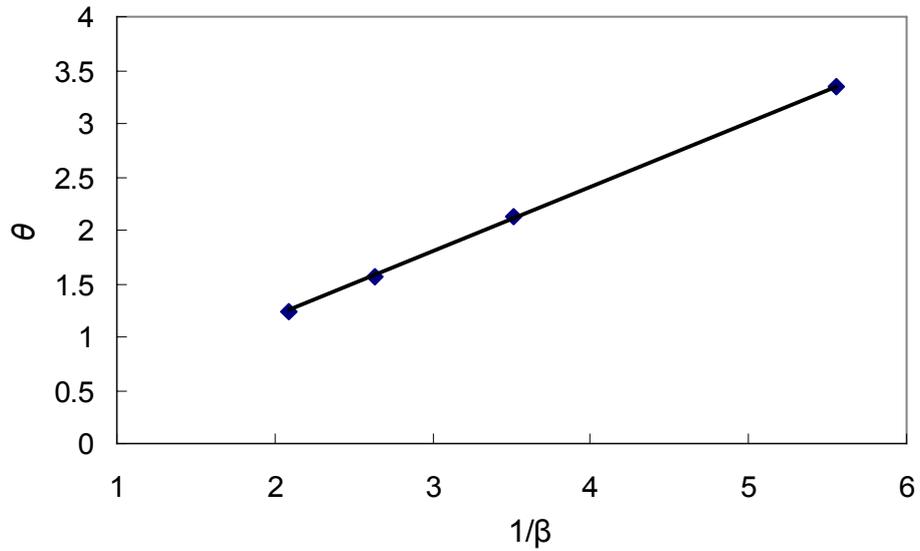

***Figure 18***.   *The $\theta$ parameter of Eq. (2) as a function of $1/\beta$.*

In the full configuration with the coupling cell, it was possible to obtain the resonant frequency of the working mode and those of its neighbors for different number *m* of gaps. In Fig. 19 the results are compared to the configuration with a single accelerating section ($N_{cc} = 0$; where $N_{cc}$ is the number of coupling cells in the structure).

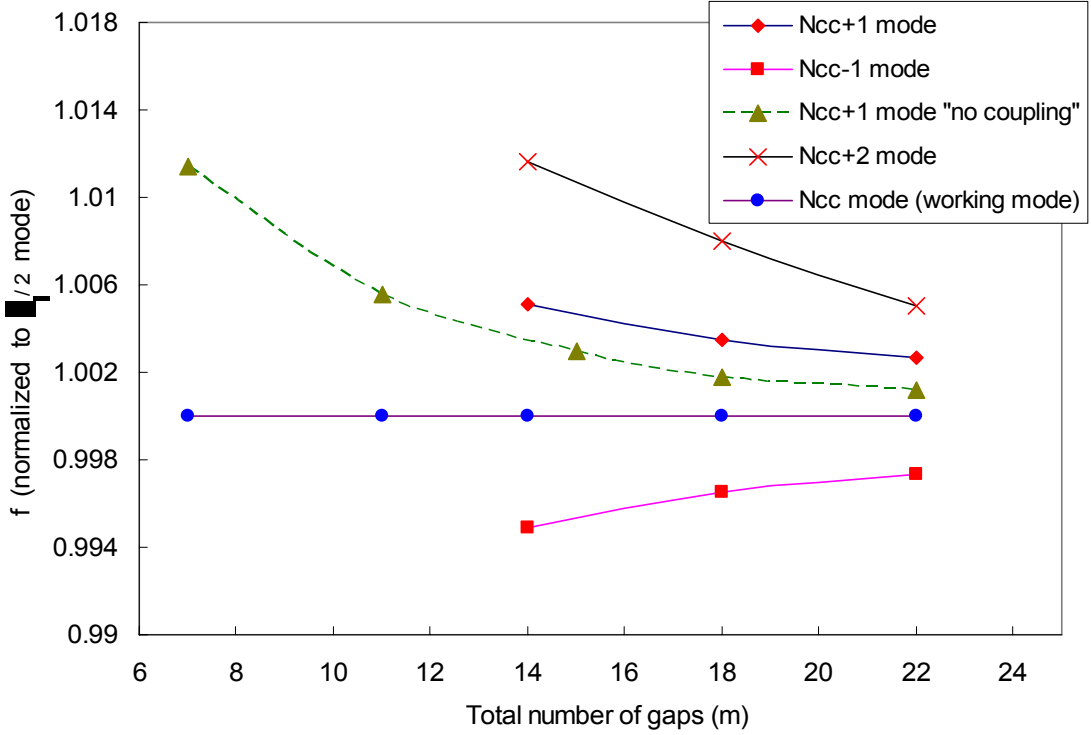

*Figure 19.* Measured frequencies for different geometrical configurations. $N_{cc}$ indicates the number of coupling cells: $N_{cc} = 0$ (dotted line) and $N_{cc} = 1$.

The working mode (points in the plot) keeps the same frequency with and without coupling cell for any number of gaps. The first superior mode without coupling cell ($N_{cc}+1$ mode with $N_{cc} = 0$; dotted line in the plot) follows Eq. (2).

By increasing the number of gaps, and consequently the length of the structure, this mode gets closer to the working mode and therefore instability increases. Fig. 19 shows the measurements of the first three resonances relative to this mode, (m = 7, 11, 15). The other two (m = 18, 22) have been extrapolated from Eq. (2) with the proportionality coefficient determined by the linear fitting of Fig. 17. When the coupling cell is introduced, the structure can be considered as a system of three coupled resonators and the working mode assumes the configuration of a $\pi/2$ mode ($N_{cc}$ mode with $N_{cc} = 1$). The frequency of this mode is the same with and without coupling cell since, for a $\pi/2$ mode, it does not have stored energy in the central coupling cell. The two neighbor modes are the $\pi$ mode ($N_{cc}-1$ mode with $N_{cc} = 1$; squares in the plot), that is the lowest in frequency, and the zero mode ($N_{cc}+1$ mode with $N_{cc} = 1$; diamonds in the plot). Both modes are dependent on the frequency of the coupling cell, whose tuning was necessary in order to close the stop band and to have a symmetric configuration of the two modes with respect to the $\pi/2$ mode.

Fig. 19 is the main result of the present paper and shows how the introduction of the coupling cell increases the frequency separation between the modes (as expected from the definition of a coupled system) and consequently it improves the stability of the system, especially for high values of *m*. These results have been obtained with small slots (four on each side of the coupling cell with an angular aperture of 40° and only 8 mm width) and small values of the coupling (Fig. 20).

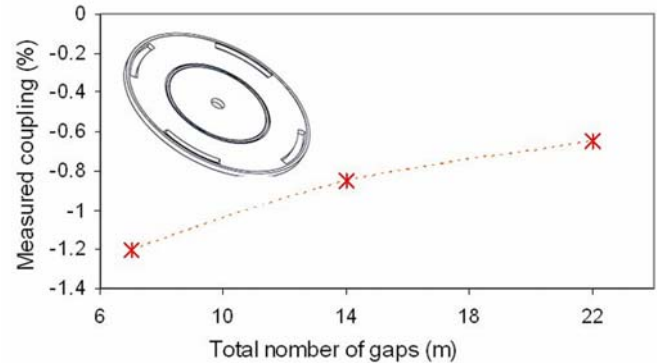

*Figure 20.* *Value of the coupling (k) for different configurations and artistic view of the end-plat with four coupling slots.*

The particular designs of both the end-cells and the coupling cell have the advantage of allowing no practical limitation in the slot dimensions: if needed, the final design could have larger slots to improve the coupling.

The measured electric field pattern along the accelerating axis is shown in Fig. 21. It has been obtained with the bead pulling technique and by tuning both the end-cells and the coupling cell, but without using any other tuning screws in the accelerating tanks.

Despite the low mechanical accuracy and the absence of precise tuning, the relative field in the different gaps is within ±3% with the exception of the four end-cells, where the field assumes a particular configuration.

Some modifications in the design of the end-cells allowed obtaining a good uniformity in the whole structure. Fig. 22 shows the field configuration in a single tank after the design of the end-cell was modified.

This simple prototype confirmed the RF field behavior as expected from the previous computations and demonstrated all the advantages provided by a bi-periodic structure working in the $\pi/2$ mode in terms of field stability. The design of both the coupling cell and the slots profited from the measurements on the prototype. The tuning procedure of the coupling cell was also successfully tested.

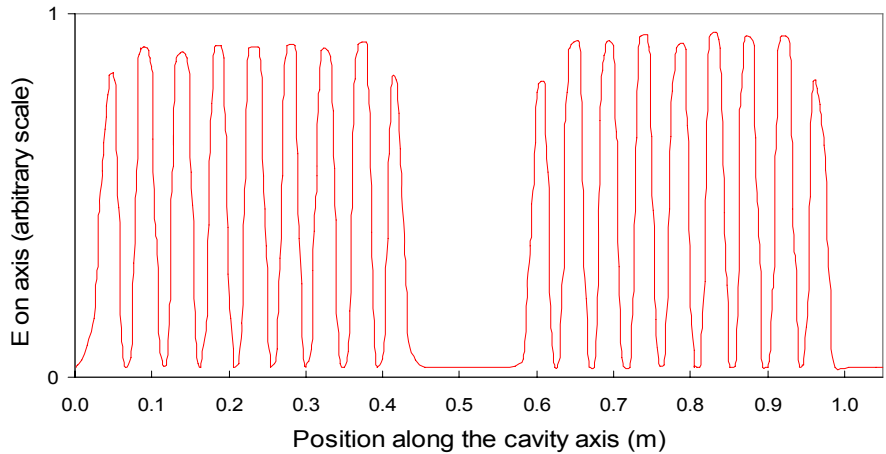

*Figure 21*.   Electric filed along the structure (absolute value) as measured by the bead-pulling technique.

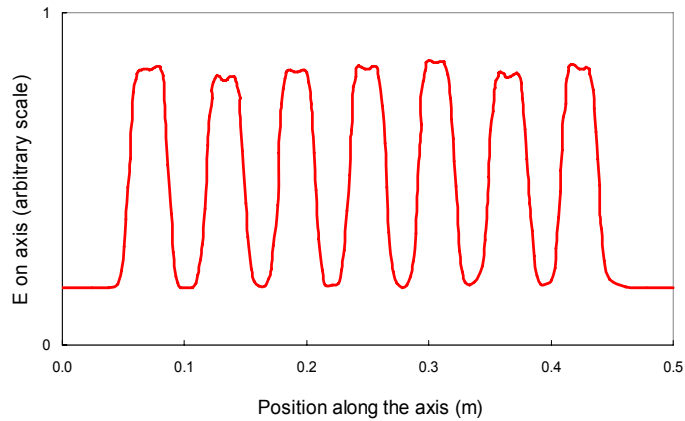

*Figure 22*.   Electric field (absolute value) in one tank after the modification of the terminal drift tubes.

## 6. DESIGN OF A 1.5 GHZ CLUSTER

In a cyclinac, CLUSTER is the natural choice to boost the proton beam, which is produced by a commercial and cheap cyclotron, up to a convenient energy for the injection into LIBO; according to the scheme of Fig. 23. A particular design of a cyclinac employing the CLUSTER structure is described in this section.

The input beam energy for the present design is 15 MeV, since all the commercial cyclotrons normally employed for FDG production are in the range 10-18 MeV [33]. This choice allows a straightforward adaptation of the design to a specific model of cyclotron.

The working frequency of CLUSTER has been reduced from 2.998 GHz (LIBO frequency) to 1.499 GHz in order to increase the beam acceptance of the DTL part of the two section linac.

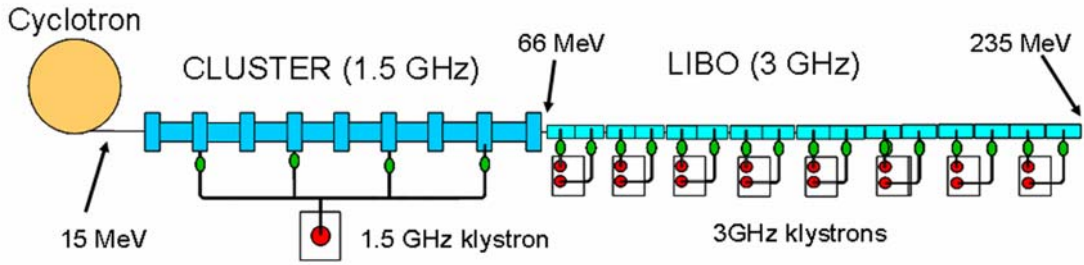

*Figure 23*.  *Scheme of IDRA based on a DTL section followed by a CCL section. The two structures are of the CLUSTER and LIBO type respectively.*

The CLUSTER-LIBO interface energy was fixed at 66 MeV which is the energy used for the treatment of eye melanomas. In the range 15-67 MeV no variation of the beam energy is required for the treatments. This gives the possibility to use a single klystron to feed four resonant cavities reducing the overall cost. The candidate klystron is the Thales TH2117 with 20 MW peak power and 14 kW average power.

The four modules of CLUSTER have been designed to require about the same power. Particular attention has been devoted to the power distribution from the single klystron to the four modules.

The basic parameters of the two sections of the linac are given in Table 2.

*Table 2 Main parameters of the linac which accelerates protons from 15 MeV to 235 MeV*.

| Type of linac | DTL CLUSTER | SCL LIBO |
|---|---|---|
| Frequency  [MHz] | 1499 | 2998 |
| Input energy  [MeV/u] | 15 | 66 |
| Output energy  [MeV/u] | 66 | 235 |
| Number of accelerating modules | 4 | 16 |
| Diameter of the beam hole  [mm] | 8 | 8 |
| Total length of the linac  [m] | 6.2 | 16.8 |
| Effective shunt impedance $ZT^2$ (inject.-extract.)  [M$\Omega$/m] | 109 - 81 | 43 - 70 |
| Average axial electric field (injection-extraction)  [MV/m] | 14.7 | 16.7-19.0 |
| Kilpatrick number defining the max. surface electric field | 2.1 | 1.8 |
| Total peak RF power for all the klystrons  [MW] | 9.0 | 54 |
| Repetition rate of the proton pulses  [Hz] | 200 | 200 |
| Pulse length  [µs] | 5.0 | 5.0 |
| Duty cycle  [%] | 0.1 | 0.1 |
| Power required by the linac (R23xR26)  [kW] | 9.0 | 54 |

This cyclinac design represents a cheaper and promising alternative to IDRA which accelerates protons starting at 30 MeV. The complete study including beam dynamics, thermo-mechanical analysis, and brazing procedure are part of the future activities of the TERA Foundation.

**Acknowledgements**
The authors would like to express their sincere gratitude to the members of the TERA Foundation, especially to Mario Weiss for his advices and expertise, to Peter Pearce for the information concerning the klystron, and to Gabriele Fagnola for the drawings of the prototype. The RF measurements benefited from the continuous help of Giulio Magrin and Saverio Braccini. Sonia Allegretti provided an important contribute to the mechanical design of the stems and the study of the thermal deformation of the structure.
A particular thank to the CERN AB/RF group and especially to Georges Carron for the RF instrumentation and the technical support.
The development of CLUSTER, from the beginning to the present design, would not have been possible without the financial support of the Monzino Foundation (Milan); A.S.P. (Associazione per lo sviluppo scientifico e tecnologico del Piemonte) (Turin) and the Price Foundation (Geneva).

**References**

[1] U. Amaldi, S. Braccini, *Present and future of hadrontherapy*, AIP Conf. Proc., 827 (2006)

[2] R.R. Wilson, *Radiological use of fast protons*, Radiology, 47 (1946)

[3] T. Kamada, H. Tsujii, H. Tsuji, T. Yanagi. J-E. Mizoe, T.Miyamoto, H. Kato, S. Yamada, S. Morita, K. Yoshikawa, S. Kandatsu, A. Tateishi, *Efficacy and safety of carbon ion radiotherapy in bone and soft tissue sarcomas*, J. Clin. Oncol., 20 (2002)

[4] D. Schulz-Ertner, A. Nikoghosyan, C. Thilmann, T. Haberer, O. Jäkel, C. Karger, G. Kraft, M. Wannenmacher, J. Debus, *Results of carbon ion radiotherapy in 152 patients*, Int. J. Radiat. Biol. Phys., 58 (2004)

[5] U. Amaldi, G. Kraft, *Radiotherapy with beams of carbon ions*, Rep. Prog. Phys., 68 (2005); U. Amaldi, G. Kraft, *Recent applications of synchrotrons in cancer therapy with carbon ions*, Europhysics News, 36/4 (2005)

[6] Y. Jongen, W. Kleeven, S. Zaremba, D. Vandeplassche, W. Beeckman, V.S. Aleksandrov, G.A. Karamysheva, N.Yu. Kazarinov, I.N. Kian, S.A. Kostromin, N.A. Morozov, E.V. Samsonov, G.D. Shirkov, V.F. Shevtsov, E.M. Syresin, *Design studies of the compact superconducting cyclotron for hadron therapy*, EPAC Conf. Proc., Edinburgh, Scotland (2006)

[7] E. Pedroni, R. Bacher, H. Blattmann, T. Böhringer, A. Coray, A. Lomax, S. Lin, G. Munkel, S. Scheib, U. Schneider, A. Tourosvsky, *The 200-MeV proton therapy project at the Paul Scherrer Institute: conceptual design and practical realization*, Med. Phys., 22 (1995)

[8] U. Amaldi, M. Grandolfo, L. Picardi (Editors), *The RITA network and the design of compact proton accelerators*, Ch. 4 ; INFN - Laboratori Nazionali di Frascati (1996)

[9]  P. Berra, S. Mathot, E. Rosso, B. Szeless, M. Vretenar, U. Amaldi, K. Crandall, D. Toet, M.Weiss, R. Zennaro, C. Cicardi, D. Giove, C. De Martinis, D. Davino, M. R. Masullo, V. G. Vaccaro, *Study, construction and test of a 3 GHz proton linac-booster (LIBO) for cancer therapy*, EPAC Conf. Proc., (2000) Wien, Austria

[10] B. Szeless, P. Berra, E. Rosso, M. Vretenar, U. Amaldi, K.R. Crandall, D. Toet, M. Weiss, R. Zennaro, C. Cicardi, D. Giove, C. De Martinis, D. Davino, M.R. Masullo, V.G. Vaccaro, *Successful high power test of a proton linac booster (LIBO) prototype for hadrontherapy,* PAC Conf. Proc., (2001) Chicago, IL, USA

[11] C. De Martinis, C. Birattari, D. Giove, L. Serafini, P. Berra, E. Rosso, B. Szeless, U. Amaldi, K. Crandall, M. Mauri, D. Toet, M.Weiss, R. Zennaro, M. R. Masullo, V. G. Vaccaro, L. Calabretta, A. Rovelli, *Beam tests on a proton linac booster for hadrontherapy*, EPAC Conf. Proc., (2002) Paris, France

[12] U. Amaldi, P. Berra, K. Crandall, D. Toet, M. Weiss, R. Zennaro, E. Rosso, B. Szeless, M. Vretenar, C. Cicardi, C. De Martinis, D. Giove, D. Davino, M.R. Masullo, V. Vaccaro, *LIBO - a linac-booster for protontherapy: construction and tests of a prototype*, Nucl. Instr. Meth. A, 521 (2004)

[13] J.P. Blewett, *Linear accelerator injector for proton synchrotrons*, Proc. CERN Symposium in High-Energy Accelerators and Pion Physics, (1956) Geneva, Switzerland

[14] P.M. Zeidlits, V.A. Yamnitskii, *Accelerating systems employing H-type waves,* J. Nucl. Energy, Part C Plasma Phys., 4 (1962)

[15] U. Ratzinger, *The new GSI prestripper linac for high current heavy ion beams*, LINAC 96 Conf. Proc., (1996) CERN, Geneva, Switzerland

[16] N. Angert, W. Bleuel, H. Gaiser, G. Hutter, E. Malwitz, R. Popescu, M. Rau, U. Ratzinger, Y. Bylinski, H. Haseroth, H. Kugler, R. Scrivens, E. Tanke, D. Warner, *The IH linac of the CERN lead injector*, LINAC 94 Conf. Proc., (1994) Tsukuba, Japan

[17] R. Zennaro, *Studio della stabilizzazione del campo accelerante in un RFQ di 5 MeV tramite la realizzazione di un modello in scala reale* (in Italian), Laurea Thesis (1998), Università di Ferrara, Italy

[18] J.H. Billen, F.L. Krawczyk, R.L. Wood, L.M. Young, *A New RF structure for intermediate-velocity particles*, LINAC 94 Conf Proc., 1 (1994), Tsukuba, Japan

[19] L. Picardi, *RF behaviour of 3 GHz SCDTL structures*, Eur. Phys. J. AP 20, (2002)

[20] D.E. Nagle, E.A. Knapp and B.C. Knapp, *Coupled resonator model for standing wave accelerator tanks*, Rev. Sci. Instr., 38 (1967)

[21] J. Gao, *Analytical formulas for the resonant frequency changes due to opening apertures on cavity walls*, Nucl. Instr. Meth. A, 311 (1992)

[22] T. Weiland, *Solving Maxwell's equations by means of the MAFIA-CAD system*, DESY report, (1988), Hambourg, Germany

[23] http://www.cst.com

[24] U. Ratzinger, *H-Type linac structures*, CERN Accelerator School (2000), Seeiheim, Germany

[25] R. Zennaro, *IDRA: design study of a protontherapy facility*, ICFA Beam Dynamics Newsletter, 36 (2005)

[26] C. Cianfarani, E. Cisbani, G. Orlandi, S. Frullani, L. Picardi, C. Ronsivalle, *Status of the TOP linac project*, Nucl. Instr. Meth. A, 562 (2006)

[27] S.O. Schriber, *Recognizing, analyzing and interpreting mode information in standing-wave cavities*, LINAC 2002 Conf. Proc., (2002) Gyeongju, Korea

[28] C. Tronci, *Analisi del campo elettromagnetico a radiofrequenza nel progetto di un linac di nuova concezione per l'adroterapia con ioni carbonio* (in Italian), Laurea Thesis (2004), Politecnico di Torino, Italy

[29] S.Y. Lee, *Accelerator Physics*, Ch. 8, World Scientific, 1999

[30] S.O. Schriber, *Analog analysis of $\pi$-mode structures: Results and implications*, Phys. Rev. ST Accel. Beams, 4 (2001)

[31] T.P. Wangler, *Introduction to linear accelerators*, Los Alamos National Laboratory report LA-UR-94-125 (1994), Los Alamos, NM, USA

[32] P.M. Lapostolle, M. Weiss, *Formulae and procedures useful for the design of linear accelerators*, CERN report CERN-PS-2000-001-DR (2000), Geneva, Switzerland

[33] http://ribfweb1.riken.go.jp/cyc2004/proceedings/data/ListOfCyclotrons.html